# WHY MONOTONOUS REPETITION IS UNSATISFYING

*Nikos A. Salingaros, The University of Texas at San Antonio.*

**Abstract**[1]: *Human beings prefer ordered complexity and not randomness in their environment, a result of our perceptual system evolving to interpret natural forms. We also recognize monotonously repeating forms as unnatural. Although widespread in today's built environment, such forms generate reactions ranging from boredom to unease. Christopher Alexander has introduced rules for generating forms adapted to natural geometries, which show structured variation with multiple symmetries in a hierarchy of scales. It turns out to be impossible to generate monotonously repeating forms by following those rules. As it is highly probable that traditional artifacts, buildings, and cities were created instinctively using a version of the same rules, this is the reason we never find monotonously repeating forms in traditional cultures.*

**1 Conjectures on combinatorial complexity.**

When applying mathematics to interpret our world we invariably run into formidable difficulties. Explaining human perception of our surroundings and our reactions to the environment requires that we know the mechanisms of our interaction with the world. Unfortunately, we don't — not yet. Thus, explanations of why we react to different forms in our environment tend to be conjectural.

We know from observation that human beings crave structured variation and complex spatial rhythms around them, but not randomness. Monotonous regularity is perceived as alien, with reactions ranging from boredom to alarm. Traditional architecture focuses on producing structured variation within a multiplicity of symmetries. Contemporary architecture, on the other hand, advocates and builds structures at those two extremes: either random forms, or monotonously repetitive ones. Let us explore why human beings find the latter unappealing, and propose what they do like instead, with the ultimate aim of characterizing that mathematically.

I present some ideas on design and the influence of certain structures on human perception. These questions arose in the context of architecture and urbanism, yet the problem goes much deeper, into combinatorics and human physiological response.

---

[1] **Keywords**: Complexity, patterns, symmetry, symmetry breaking, architecture, perception, Biophilia, Christopher Alexander, structured variation, scaling hierarchy, mathematical aesthetics.

Lacking a rigorous theory that explains the phenomenon, I am offering some intuitive results in the hope that someone will pursue them further. I believe there exists a straightforward mathematical model for what is going on.

The observed effect concerns modules repeating in a geometrically regular manner: an application of translational symmetry. For example, consider identical rectangles lined up straight. On the scale of skyscrapers, many of those buildings simply repeat the floor design (as seen on its side) vertically, so that the whole building shows vertical translation symmetry (Figure 1).

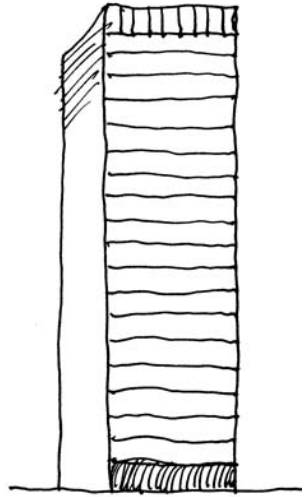

*Figure 1. A skyscraper shows vertical repetition.*

There are also countless examples of exact repetition of units horizontally, either identical structural elements within one building's façade (Figure 2), or separate but identical buildings lined up straight along a road. A typical example is the repeating modular box making up an urban housing or office development (Figure 3).

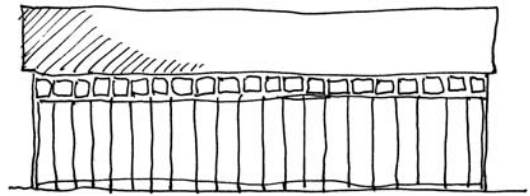

*Figure 2. Non-traditional building showing horizontal repetition.*

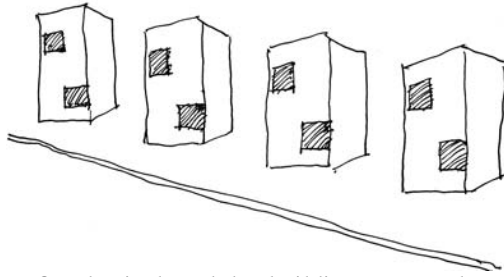

*Figure 3. Identical modular buildings repeat along a street.*

Many observers react negatively to such simplistic repetition. And, apparently, our degree of unease increases as more units are seen to repeat. Identical objects perfectly lined up generate a feeling of discomfort in the viewer. At the very least, we experience something mechanical to which we react negatively as human beings, since we are used to more natural structures with variation and complexity. Readers are encouraged to check this assertion. Observations should be performed with the entire body in a full-scale environment: simply looking at pictures or at a reduced-scale model fails to engage all of our perceptive apparatus and will not lead to useful results. The conjecture is that something fundamental is at play here that affects our perceptive mechanism, triggering a negative signal.

So, what do people prefer? And why? I open the can of worms of architectural fashion by illustrating an older example of tall building typology, dating from the end of the 19th to the beginning of the 20th centuries (Figure 4). Monotonous vertical repetition is absent. In addition, a richness of articulation that is part of traditional tectonics and ornamentation provides a hierarchy of decreasing scales.

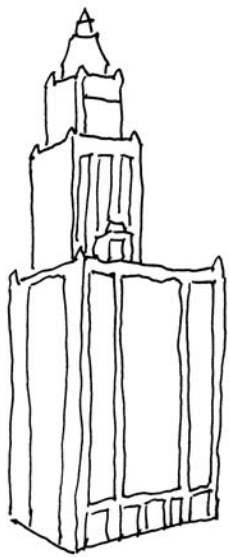

*Figure 4. Neo-Gothic skyscraper.*

Even though architects since the 1920s advocate monotonous repetition in tall buildings (Figure 1), it is not really appealing to most people. I believe buildings with more complex yet ordered shapes make a far better city — but then, they must also pay attention to the human scale (which is another topic for discussion).

**2 Steps toward an explanation.**

First, the mathematics points us to the algorithmic complexity of the configuration. The simplest complexity measure considers the generating code of a configuration. Complexity can be measured as being directly proportional to the length of code (though this is in itself a simplistic measure that does not apply in more sophisticated situations, it is sufficient here). In this case, repetition in one direction is trivially simple, since the code for generating it is: "define a unit A, then align $n$ identical copies along one direction to get AAAAAAAA… ". The generating code is trivial.

Second, why is human neurological response actually negative? Some insight into the effect comes from the notion of Biophilia, which asserts that our evolution formed our neurological system within environments defined by a very high measure of a specific type of coherent complexity. That is, our neurological system was created (evolved) to respond directly and exquisitely to complex, fractal, hierarchical geometric environments. When placed in environments that have opposite geometrical features, therefore, we feel ill at ease. The theory is being verified by experiments: see the collection of essays "Biophilic Design" edited by Stephen Kellert *et al*. Minimalist environments make us feel ill at ease. Simplistic repetition is one such minimalistic geometrical setting in which we find no algorithmic complexity, hence no visual and intellectual interest. But our response is not simply to ignore it: we cannot, and it provokes anxiety in us.

Questions immediately arise, such as why does the feeling of unease increase with the number of repetitions? Here is evidently a straightforward Combinatoric problem, if only we knew what the human brain is measuring or counting when looking at repeating modules. But we don't. One guess is that the feeling of unease grows not linearly but exponentially with the number $n$ of repeating modules. If the brain is counting permutations among identical units, or trying to label each unit, then the possible combinations increase exponentially. It could also be true that the brain is frustrated by trying to identify distinct modular units so as to fix a coherent picture of its environment — necessary for survival and deciding upon a fight-or-flight response. If the units are identical they cannot be catalogued in memory.

**3 Some examples.**

In what environments does one encounter large-scale geometrical configurations with a lot of monotonous repetition? Actually, all such examples are human-made, being strictly the results of industrial production. I claim that simplistic repetition occurs neither in nature, nor in pre-industrial human creations.

In nature, we almost never find simplistic repetition on the macroscopic scale. (Yes, pure crystals do have microscopic regularity but that structure is not visible — furthermore, naturally-occurring pure crystals are quite rare.) Inanimate physical structures almost always have some variations that prevent the unpleasant monotonous effect. For example, the most regular repetition I can immediately think of occurs in the hexagonal cells of honeybees and solidified lava flows (the Giant's Causeway); but in each of those cases the repetition occurs with abundant minor imperfections (Figure 5). Those geometries therefore avoid the "industrial" feeling of being monotonous. Looking at wild honeycombs is fascinating, and walking on crystallized lava is as well. And neither of these structures is a common part of our living environment. Other physical geometries defined by repetitions occur with a great deal of variation and so escape monotony.

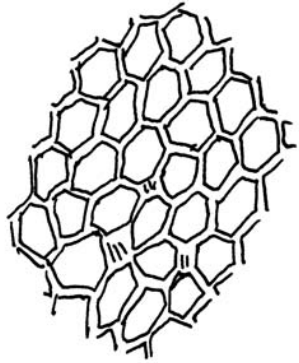

*Figure 5. Natural honeycomb.*

Living structures with repetition show so much variation in the repetition that monotony is entirely avoided. Consider the leaves of a tree: no two are identical, and their positioning combines a distribution based on the Fibonacci sequence with randomness due to the growth of the tree branches as influenced by environmental factors. No simplistic repetition along a line here!

Large-scale hexagons have been used in buildings. Where an additional variation is introduced to distinguish the modules, the result succeeds, but where the modules repeat monotonously, the overall effect is felt negatively. The architects of those buildings appear unaware of the effect of monotonous repetition, but then so many buildings from the past one hundred years blatantly display monotonous repetition, so it seems to be something desired rather than arrived at accidentally. As the experience is not yet rigorously documented it could be dismissed as personal opinion or preference. Nevertheless, I believe this is NOT personal preference but instead the reaction of our bodies, and is thus felt by the general population.

**4 Avoiding combinatorial complexity.**

I introduced the notion of combinatorial complexity in the book "Twelve Lectures on Architecture". This is precisely the effect of monotonous repetition experienced in the environment. Two solutions were given of how to avoid combinatorial complexity. Both solutions involve breaking the translational symmetry in some way.

The first solution is to introduce symmetry breaking by means of variety in the repeating modules (Figure 6). Symmetry breaking is a key notion that comes from theoretical physics: one adds small differences to an otherwise perfect symmetry. The configuration is NOT symmetric unless we ignore those minor differences. Therefore, there is approximate symmetry on the global scale but it does not extend to the smaller scales. In the present example, we maintain the translational symmetry on the larger scale (the repetition of a modular unit), but introduce variations within each module so every module is only approximately the same. Strictly speaking, it's no longer a module. These small variations are sufficient to affect our perception of the whole configuration, however, changing it from being monotonous to interesting.

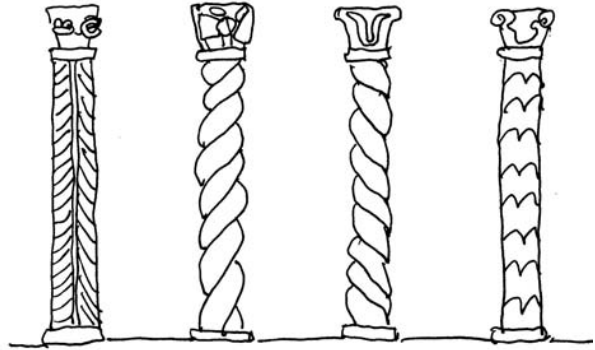

*Figure 6. Columns with variety, spaced symmetrically.*

The second solution is to group a few modules together into a cluster of no more than three or four (Figures 7 and 8). We somehow tie three or four modules into a supermodule, which itself then repeats. What we are doing is in fact introducing a hierarchy where previously none existed. In the original repeating configuration of *n* modular units, the scales are only two: the module itself, and all the modules lined up filling the size of the entire configuration. By grouping modules, we define a new scale at the size of the grouping, not exactly three or four modules large, because it includes modules plus any intermediate spaces.

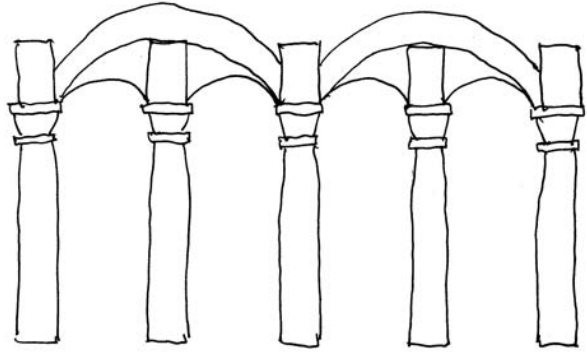

*Figure 7. Grouping columns into clusters of three.*

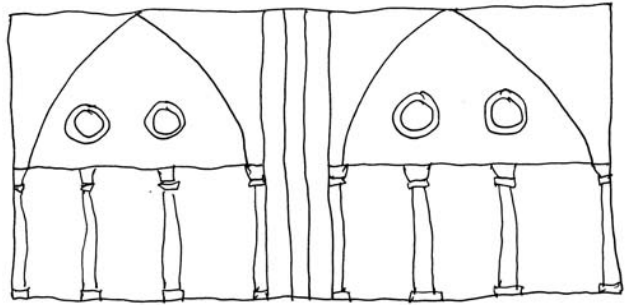

*Figure 8. Grouping columns into clusters of four.*

These two groupings establish a strong informational relation between what we initially considered to be the module and any space surrounding it; grouping columns with spaces links alternating contrasting units. Rhythm on both the architectural and urban scales depends upon the intricate interweaving of space and material, which are treated on an equal design footing.

It is worth pointing out that these solutions come from traditional architecture, and are seen to be re-invented repeatedly by different cultures in history and in different geographical regions. Something innate is driving humankind towards discovering and implementing these solutions, and it's not simply a matter of aesthetic preference. Also note that our modern industrial age (beginning, say, from the 1920s) is marked by its break with the architecture of the past by the distinction of whether to pursue and celebrate monotonous repetition, versus avoiding it altogether. Since the effect produces unease in the user, this raises serious questions about why architects and urbanists make it a point to generate it in their buildings.

**5 Christopher Alexander's explanations.**

Christopher Alexander is a pioneer in investigating environmental complexity and developing techniques for generating living geometry in the built environment. By "living geometry" we mean a particular complexity that embodies coherence and which is perceived as physiologically and psychologically positive by human beings. Alexander refers to the environmental effect as "healing", confirming the independent line of investigation coming from Biophilia. Although not expressed in the present manner, Alexander's work offers fundamental insight into the problem of monotonous repetition.

Alexander has presented "Fifteen Fundamental Properties" that are found in all coherent structures, comprising inanimate matter, biological structures, and especially in human-made objects and environments built before the industrial age. Those rules can be applied to generate living environments today, and are clearly useful in avoiding, or repairing, monotonous repetition so as to remove its negative effect. A reader can find descriptions of these properties in Alexander's "The Nature of Order. Book 1: The Phenomenon of Life", and my own summary in "Twelve Lectures on Architecture". I describe three of Alexander's properties here.

1. ***Levels of Scale*** postulates that stable structures contain a hierarchy of distinct scales, and those scales are carefully spaced so that the scaling factor between two consecutive scales is very roughly equal to three. This is a universal property satisfied, for example, by all fractals. (Whereas fractals exist with every scaling factor, Alexander postulates that hierarchies with scaling factor near three are perceived as more natural). There should exist distinct scales well defined in the structure. The larger scales are related through some magnification (exact or approximate) to the smaller scales, using a scaling factor. As described above, grouping repeating units into clusters introduces intermediate scales where none existed initially. As such, it is one solution to avoiding or repairing monotonous repetition.

For example, what makes a colonnade informationally comfortable depends just as much on hierarchy as on repetition. Inter-columnar spacing ranges from two column widths in some Classical temples, to four in the nave arcade of a Basilica and in Roman colonnades, to six in many Medieval Cloisters, to eight in Far Eastern traditional architecture (with variations for individual cases). In all these instances, the space between columns defines the next higher scale, and repetition links two consecutive hierarchical scales. As a result, the columns and spaces are perceived together coherently. Twentieth-century architects introduced extremely thin columns (called "pilotis" or stilts) and widened their separation so the intervening space is more than twelve times the width of a column.

2. ***Alternating Repetition*** postulates that simple modules should not repeat, but rather, it is paired contrasting modules that can do so. There are several consequences of this rule. The alternation of units leads to contrast, which introduces spatial rhythm (albeit primitive but at least present, whereas monotonous repetition has no spatial rhythm at all) (Figure 9). Looking at natural, biological, and pre-industrial structures Alexander found alternating repetition to be widespread, and noted that it was adopted as a technique for creating stable configurations.

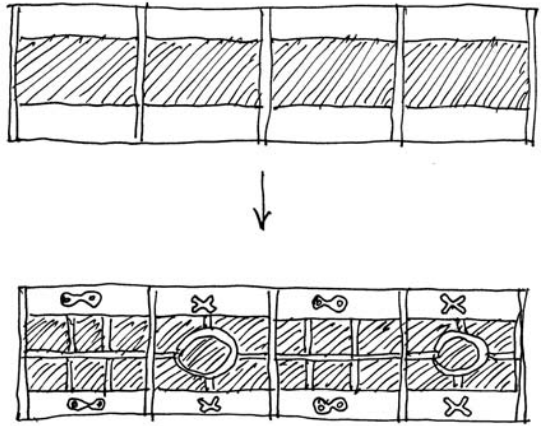

*Figure 9. With smaller-scale insertions, windows become alternating.*

Alternating repetition is directly related to the groupings discussed earlier. An alternating pattern ABABABABA… really includes symmetric groupings on many distinct scales: ABA, BAB, ABABA, BABAB, etc. defined naturally through their bilateral symmetry. This wealth of subsymmetries is not evident in configurations with monotonous repetition. For a discussion, see "The Nature of Order. Book 1: The Phenomenon of Life". Even so, a configuration with alternating repetition but no further variations or groupings on higher scales will show monotonous repetition if it is large enough.

3. **_Gradients_** occur when the size of similar components decreases in one direction. Here is a solution that was not mentioned earlier in this essay, and which breaks monotonous repetition: make all the units of different size, not randomly, but in a carefully controlled manner so as to create a gradient. Then, the ensemble is perceived as harmonious and not unsettling. Translational symmetry is broken because the units have decreasing lengths. Again, Alexander found countless examples of gradients in natural, biological, and pre-industrial structures.

Gradients prevent monotonous repetition, but in a different manner to symmetry breaking. In the latter, the main symmetry is maintained while symmetry is broken on smaller scales. Gradients, on the other hand, break the symmetry on the original scale and do not necessarily do anything on smaller scales.

Here I mentioned only three of Alexander's fifteen fundamental properties, yet there are other ones that bear directly and indirectly upon our problem of monotonous repetition. All of this discussion is but a small part of a general theory of design that uses recursive algorithms [a good topic for a future paper in this series]. Assuming as axiomatic several geometrical properties found in nature, Alexander has formulated a method for designing and constructing complex systems.

Interestingly, following Alexander's design method "The Theory of Centers", one cannot get to monotonous repetition. It's not that monotonous repetition is forbidden in any ideological sense; rather the algorithmic design rules can never arrive at solutions that display monotonous repetition. That mechanical configuration, and most other unnatural, anxiety-inducing geometries, resides outside the space of solutions obtainable from the Theory of Centers. To reach monotonous repetition, one has to abandon the design rules that generate living structure. Therefore, since the Theory of Centers was most certainly followed by designers and builders of all traditional cultures instinctively, this explains why we never see monotonous repetition in traditional artifacts, buildings, and cities.

### 6 Symmetry breaking prevents informational collapse.

An identically repeating module generates simplistic translational symmetry. As explained previously, such a configuration has algorithmically trivial complexity. From the information theory point of view, the configuration is collapsible to its single module plus the rule for repetition. Thus, the configuration as a whole has no informational stability: it is prone to collapse. Symmetry breaking changes this because it is no longer possible to condense the whole into a single repeating module.

There may be more to it than simple visual concerns about monotony in design. A physical structure is only one of an enormous variety of complex systems that run our universe. Each complex structure must protect itself against structural collapse, otherwise we will not find it around to observe. Does informational collapse parallel other, more significant mechanisms of systemic collapse? And do complex systems find analogous methods of avoiding systemic collapse?

Since symmetry breaking through the creation of higher-order groupings generates a hierarchy, this itself is one basic feature of a stable system. By introducing distinct levels in a scaling hierarchy, the complex system distributes itself on different levels, and thus it is not dependent solely upon one or two levels. Symmetry breaking as seen in nature and in traditional artifacts, buildings, and cities is not random, but serves to define an irreducible hierarchical structure. This question is discussed further in "Twelve Lectures on Architecture".

### 7 Blending into the background and emotional nourishment.

Natural environments are characterized by an enormous degree of structural complexity, yet for the most part, we perceive them as background. Our perceptive system is apparently wired to notice anything that contrasts with a natural background. It signals alarm and makes us uneasy. Since natural environments are fractal, it follows that non-fractal objects will stick out and be noticed by us. This includes pure Platonic forms (cubes, rectangular prisms, pyramids, spheres) that define just a single scale, the largest one. Usually, those repeating forms create the monotonous repetition effect discussed here.

Andrew Crompton sent me some helpful suggestions on the topic of this essay: Human creations are designed either as neutral or picturesque, and traditional products are designed to vanish into our surroundings. When we inhabit an environment, or surround ourselves with human-made objects, we don't want any individual object to bother us — that is, not to disrupt the sense of visual coherence we can draw from our complex environment. Repetitive buildings or building components may be computationally boring, but they do impose themselves upon our cognition by not blending into the background. They stick out. They do not scale (as I discussed earlier) so they do not fit into a traditional structural hierarchy of a city that has evolved over generations.

Monotonous repetition disturbs us because it is unnatural; and is so because it fails to share geometrical features common in natural complex structures. The phenomenon goes further, however, in that "blending into the background" is not a neutral effect, but definitely a regenerative one. Biophilia gives us emotional nourishment (with concomitant physiological benefits) from a complex, coherent environment; therefore I am talking about positive effects.

**8 Thoughts about contemporary architecture.**

In contemporary architecture, many practitioners have rebelled against monotonous repetition and have come up with their own solutions. Invariably, those solutions inject randomness into the translational symmetry in a way that leads away from coherence. This is the opposite from the solutions outlined above and implemented by traditional architecture and urbanism, which seek coherence. Someone who is familiar with contemporary architects' philosophy of wishing to break with the past at all costs should not be surprised that traditional evolved solutions are not adopted, but that the opposite effect is sought.

The façades and plans of many contemporary buildings rely on modular units that are for the most part monotonously repetitive. The translational symmetry is sometimes broken by random changes, however, so that the overall effect is one of imbalance, irrationality, and lack of purpose (Figure 10). This negative impression is justified since the architect simply introduces random changes for visual effect, not for any structural or functional reason. The reaction of the user is not positive, because our body also reacts to randomness with alarm.

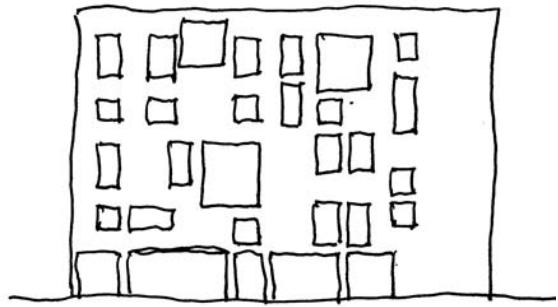

*Figure 10. Randomness in a façade destroys coherence.*

While the topic of contemporary design lies outside the present investigation, those examples of fashionable architecture that randomize symmetry contrast sharply with the solutions described here. In general, architectural symmetry breaking as practiced today violates perfect symmetries (which could be a good thing) but on the wrong scale, so that the coherence of the ensemble is reduced instead of enhanced (Figure 10).

Again the discussion goes back to our body's predilection for coherent complexity in our environment, and our negative reaction when built forms deny it to us. In its search for design novelty, architectural symmetry breaking as seen in contemporary structures deliberately avoids creating the sought-for hierarchy of subsymmetries on distinct scales.

## 9 Conclusions.

I claimed a visual effect of monotonous repetition and suggested that it induces unease and even anxiety in viewers experiencing such a structure at full scale. Hopefully, researchers in environmental psychology will perform the necessary rigorous testing in order to establish any effect such structures have on our psychology and physiology. Looking for an explanation of this effect from mathematics led me to conjecture some sort of combinatorial analysis that our brain engages in, the details of which are as yet unknown. The process of analyzing our environment occurs automatically because we need to position our bodies within it informationally, and subconsciously judge our safety from environmental threats. If an environment embodies monotonous repetition, it could tire our neurological system, and that is possibly what creates a negative effect on our bodies. This essay concluded with suggestions for avoiding the effect of monotonous repetition. Altogether, I believe this is a pretty but not well-defined, hence woefully under-investigated mathematical problem.

(Figures drawn by Nikos A. Salingaros)